\documentclass[aps,prd,reprint,preprintnumbers,showpacs,%
superscriptaddress,nofootinbib,amsmath,amssymb]{revtex4-1}

\begin{document}

\preprint{arXiv:1307.6534 [gr-qc]}

\title{Conformal Stealth for any Standard Cosmology}

\author{Eloy Ay\'on--Beato}
\email{ayon-beato-at-fis.cinvestav.mx}
\affiliation{Departamento~de~F\'{\i}sica,~CINVESTAV--IPN,%
~Apdo.~Postal~14--740,~07000,~M\'exico~D.F.,~M\'exico.}

\author{Alberto A. Garc\'{\i}a}
\email{aagarcia-at-fis.cinvestav.mx}
\affiliation{Departamento~de~F\'{\i}sica,~CINVESTAV--IPN,%
~Apdo.~Postal~14--740,~07000,~M\'exico~D.F.,~M\'exico.}
\affiliation{Departamento~de~F\'{\i}sica,~UAM--I,~Apdo.~Postal~55--534,~09340,~M\'exico~D.F.,~M\'exico.}

\author{P. Isaac Ram\'irez--Baca}
\email{pramirez-at-fis.cinvestav.mx}
\affiliation{Departamento~de~F\'{\i}sica,~CINVESTAV--IPN,%
~Apdo.~Postal~14--740,~07000,~M\'exico~D.F.,~M\'exico.}

\author{C\'esar A. Terrero--Escalante}
\email{cterrero-at-ucol.mx}
\affiliation{Facultad de Ciencias,~Universidad de Colima,%
~C.P.~28045.~Colima,~Col.,~M\'exico.}

\begin{abstract}
It is shown that \emph{any} homogeneous and isotropic universe,
independently of its spatial topology and matter content,
allows for the presence of a conformal stealth, i.e.\ a
nontrivial conformally invariant scalar field with vanishing
energy-momentum tensor, which evolves along with the universe
without causing even the smallest backreaction. Surprisingly,
this gravitationally invisible universal witness is
inhomogeneous with zero consequences for the underlying
cosmology. Additionally, it is shown that these results are not
exclusive of a four-dimensional universe by generalizing them
to higher dimensions.
\end{abstract}


\maketitle

\section{Introduction\label{sec:intro}}

Today cosmologists are living exciting times because of the
continuous arrival of data from highly accurate satellite and
ground-based observations (see for instance
Refs.~\cite{Bernabei:2009zzd,
Suzuki:2011hu,Schelgel:2011zz,Hinshaw:2012fq,Ahn:2012fh,Ade:2013lta}).
These data are already able to tightly constrain the
theoretical description of the evolution of our universe,
arguably pointing out the so-called $\Lambda CDM$ model as the
best framework for this description \cite{Ade:2013lta}. This
model is a realization of the standard cosmology, i.e.\ a
homogeneous and isotropic universe
\begin{equation}\label{eq:FRW}
ds^2=a(\tau)^2\left(-d\tau^2+\frac{dr^2}{1-kr^2}
+r^2\left(d\theta^2+\sin^2\!\theta\,{d}\phi^2\right)\right),
\end{equation}
which for a given matter content $\varphi_{\text{m}}$
extremizes the action
\begin{equation}\label{eq:action}
S[g,\varphi_{\text{m}}]=\int{d}^4x\sqrt{-g}\left(
\frac1{2\kappa}(R-2\Lambda)+L_{\text{m}}\right),
\end{equation}
by solving the Einstein equations
\begin{equation}\label{eq:EEqs}
G_{\mu\nu}+\Lambda g_{\mu\nu}-\kappa{T}_{\mu\nu}=0.
\end{equation}

As is common practice in General Relativity, in this framework
it is always considered that any component of the matter
content will necessarily leaves a trace in the spacetime
geometry. Nevertheless, there are special nontrivial matter
configurations with no backreaction on the gravitational field.
Scalar fields with this property have been found for the static Ba\~nados--Teitelboim--Zanelli (BTZ) black hole \cite{AyonBeato:2004ig}, Minkowski flat space
\cite{AyonBeato:2005tu}, and (anti-)de Sitter [(A)dS] space
\cite{Ayon-Beato:SAdS}. They were coined gravitational
\emph{stealths}. In the cosmological context, the existence of
stealths has been shown for the de~Sitter cosmology
\cite{Banerjee:2006pr}. The non-trivial role they play in the
probability creation of these universes has been emphasized in
\cite{Maeda:2012tu}. In this letter it is shown that not only
de~Sitter universes, but \emph{any} homogeneous and isotropic
universe, without regard to its spatial topology and matter
content, allows for the presence of a conformal stealth which
evolves along with the universe without exhibiting its
gravitational fingerprints.

We will consider the general metric (\ref{eq:FRW}), without any
assumption on the allowed spatial topology ($k=0,\pm1$), and
supplement action (\ref{eq:action}) with an additional term
describing a conformally invariant self-interacting scalar
field
\begin{equation}\label{eq:action+s}
S[g,\varphi_{\text{m}}]-\frac12\int{d}^4x\sqrt{-g}\left(\partial_{\mu}\Psi\partial^{\mu}\Psi
+\frac16R\Psi^2+\lambda\Psi^4\right).\quad
\end{equation}
Our aim is not to find the whole spectrum of configurations
described by the new action, but just those critical ones with
the special property that both sides of Einstein equations
vanish independently
\begin{equation}\label{eq:EEqs+s}
0=G_{\mu\nu}+\Lambda g_{\mu\nu}-\kappa{T}_{\mu\nu}=
\kappa{T}^{\text{s}}_{\mu\nu}=0,
\end{equation}
where
\begin{eqnarray}
T^{\text{s}}_{\mu\nu}&=&\partial_{\mu}\Psi\partial_{\nu}\Psi
-\frac{1}{2}g_{\mu\nu}\left(
\partial_{\alpha}\Psi\partial^{\alpha}\Psi+\lambda\Psi^4\right)
\nonumber\\
&&{}+\frac{1}{6}\left(g_{\mu\nu}\square-\nabla_{\mu}\nabla_{\nu}
+G_{\mu\nu}\right)\Psi^2.\label{eq:Tmunu_s}
\end{eqnarray}
The vanishing of the left hand side get us back to the starting
Friedman--Robertson--Walker (FRW) universe with its corresponding
matter source. In order to prove the existence of a stealth in
any standard cosmology we need to show that imposing the
vanishing of the energy-momentum tensor (\ref{eq:Tmunu_s})
evaluated in the Friedman--Robertson--Walker background
(\ref{eq:FRW}) is compatible with a nontrivial scalar behavior:
the stealth.

In Sec. \ref{sec:SfS} we discuss how new stealth
configurations can be derived exploiting symmetries of the
stealth action, in the present case the conformal symmetry, and
highlight how in some cases this arguments can fail to give the
correct answer due to its local nature. In Sec.~\ref{sec:CwS},
the general derivation is done explicitly for the 4-dimensional
case, which is the most interesting from a cosmologist point of
view. In future work, this will be also useful as a methodology
for finding stealth configurations for more general nonminimal
couplings not necessarily of a conformal nature and for which
no obvious conformal arguments can be exploited. In the next
two Secs.~\ref{sec:flat} and \ref{sec:curved} we show the
explicit results for flat and curved universes respectively,
how they are generalized to any number of dimensions, analyze
the impact of spacetime symmetries on the corresponding
solutions, and establish the explicit comparison with the
previously discussed local conformal arguments. In the last
section we briefly summarized our conclusions.

\section{Stealths from action symmetries: local vs. global\label{sec:SfS}}

A useful way to approach the problem of the existence of
stealths is the following. The equations determining the
stealth [the right hand side of Eqs.~(\ref{eq:EEqs+s})] can be
interpreted as demanding the gravitational background to be an
extremal of the related stealth action only [the second term in
the total action (\ref{eq:action+s})]. Under this
interpretation, a \emph{full} stealth configuration, i.e.\ a
given background together with its allowed stealth, can be
linked to another potentially different \emph{full}
configuration via any symmetry transformation of the stealth
action. In particular, having a concrete example of stealth for
a conformally invariant action implies that its whole conformal
class allows also a stealth interpretation. The above argument
works well for the case of a conformal stealth on (A)dS space
since, being (A)dS conformally flat, the related configuration
\cite{Ayon-Beato:SAdS} is conformally related to that of
Minkowski flat space \cite{AyonBeato:2005tu}.

It is well-known that the Weyl tensor vanishes for
Friedman-Robertson-Walker metric (\ref{eq:FRW}), which implies
that it is also conformally flat. That is obvious for flat
universes, $k=0$, where the conformal factor is just the scale
factor $a$. It is less trivial for curved universes, $k=\pm1$,
but explicit conformal relations exist also in these cases
\cite{Lightman:1975}. Following the above described approach to
the problem, one may wonder whether we can use these conformal
maps to find the stealths for FRW spacetimes starting from the
already known configurations of Minkowski spacetime
\cite{AyonBeato:2005tu}. However, it should be emphasized that
the most general conformal transformations mapping Minkowski
spacetime into the FRW spacetimes with curved spatial topology
are only defined locally \cite{Iihoshi:2007uz}. Just the flat
case $k=0$ is globally conformal to the Minkowski spacetime.

All the above implies that we cannot ensure that the stealth
found by the corresponding conformal transformation of the one
for Minkowski spacetime will actually be well-defined in the
whole FRW spacetime. An illustrative example of how this
mapping can fail is given by the stealth of the BTZ black hole
\cite{AyonBeato:2004ig}. This background is a rotating black
hole solution of vacuum $AdS_3$ gravity \cite{Banados:1992wn}
and therefore has zero Cotton tensor, implying that is
conformally flat. On the other hand it has constant negative
curvature meaning that is locally diffeomorphic to $AdS_3$ (in
fact, it is a proper identification of $AdS_3$
\cite{Banados:1992gq}). As we mention previously (A)dS space is
conformally flat in any dimension and by this property supports
a stealth conformally related to the one of Minkowski
spacetime. Nevertheless, if one then makes use of the
diffeomorfism between the BTZ black hole and $AdS_3$ to find
the corresponding stealth over BTZ, after imposing the global
boundary condition that defines this solution, i.e.\
identifying the rotation angle by $\phi=\phi+2\pi$, the
resulting expression is found to be multivalued. Requiring it
to be single-valued imposes the vanishing of the black hole
angular momentum. Therefore, there is not a stealth
configuration for the rotating case, even if the local
conformal transformation does exist.

Taking this discussion into account, in this letter we proceed
first to search for the global stealth configurations of FRW
models by solving in general the defining stealth constraints
$T^{\text{s}}_{\mu\nu}=0$. Later, for each curvature, we
compare with the results obtained from the corresponding local
conformal transformation. Fortunately, no discrepancy is found
in the cosmological context, as opposed to the above mentioned
three-dimensional example.

\section{Cosmology with Stealths\label{sec:CwS}}

For the purpose of accomplishing our task we find useful to use
the following redefinition
\begin{equation}\label{eq:Psi2sigma}
\Psi=\frac1{a\sigma},
\end{equation}
where the function $\sigma=\sigma(x^\mu)$ inherits the full
spacetime dependence of the scalar field.

We start by writing the off-diagonal constraints determining
the stealth, i.e.\ $T^{\text{s}}_{\mu\nu}=0$ for $\mu\ne\nu$.
Let us consider first the ones involving the conformal time
\begin{equation}\label{eq:T_tau_i}
T^{\text{s}}_{\tau{i}}=\frac{1}{3a^2\sigma^3}\,\partial^2_{\tau{i}}\sigma=0,
\end{equation}
where we label the coordinates as $\{x^\mu\}=\{\tau,x^i\}$,
with $\{x^i\}=\{r,\Omega^b\}$ denoting the full spatial
coordinates and $\{\Omega^b\}=\{\theta,\phi\}$ just the angular
ones. The above equations imply that $\sigma$ is separable as a
sum of functions for the conformal time and the spatial
coordinates. Next, we consider the off-diagonal spatial
components
\begin{equation}\label{eq:T_r_b}
T^{\text{s}}_{rb}=\frac{r}{3a^2\sigma^3}\,
\partial^2_{rb}\left(\frac{\sigma}{r}\right)=0,
\end{equation}
and note that the separable spatial dependence of $\sigma$
divided by $r$, is itself separable as a sum of functions for
$r$ and the angles. The remaining off-diagonal component is
\begin{equation}\label{eq:T_theta_phi}
T^{\text{s}}_{\theta\phi}=\frac{\sin\theta}{3a^2\sigma^3}\,
\partial^2_{\theta\phi}\left(\frac{\sigma}{\sin\theta}\right)=0,
\end{equation}
that is, the angular dependence of $\sigma$ divided by
$\sin\theta$ turns out to be separable as a sum for the angles.
Summarizing, the study of the off-diagonal components leads to
the following separable form for the function $\sigma$,
\begin{equation}\label{eq:sigma_s}
\sigma(x^\mu)=T(\tau)+R(r)+r\left[\Theta(\theta)
+\sin\theta\Phi(\phi)\right],
\end{equation}
There exists a freedom in the election of the above functions,
concretely, homogeneous terms in $\Phi$, $\Theta$, and $R$ can
be compensated by a sinusoidal dependence in $\Theta$, a linear
one in $R$, and another homogeneous term in $T$, respectively.

Let us study now the diagonal components; the mixed
combinations
\begin{eqnarray}
3ra^4\sigma^3\left(T_{~\theta}^{\text{s}~\theta}
-T_{~r}^{\text{s}~r} \right) &=&
\frac{\text{d}^2\Theta}{\text{d}\theta^2}+\Theta
-(1-kr^2)r\frac{\text{d}^2R}{\text{d}r^2}\nonumber\\
&&{}+\frac{\text{d}R}{\text{d}r}=0,\nonumber
\\
3r\sin\theta{a}^4\sigma^3
\left(T_{~\phi}^{\text{s}~\phi}
-T_{~\theta}^{\text{s}~\theta}\right) &=&
\frac{\text{d}^2\Phi}{\text{d}\phi^2}+\Phi
-\sin\theta\frac{\text{d}^2\Theta}{\text{d}\theta^2}\nonumber\\
&&{}+\cos\theta\frac{\text{d}\Theta}{\text{d}\theta}=0,\nonumber
\end{eqnarray}
give rise to the following separable equations
\begin{eqnarray}
\frac{\text{d}^2\Theta}{\text{d}\theta^2}+\Theta
&=&(1-kr^2)r\frac{\text{d}^2R}{\text{d}r^2}
-\frac{\text{d}R}{\text{d}r},
\\
\frac{\text{d}^2\Phi}{\text{d}\phi^2}+\Phi
&=&\sin\theta\frac{\text{d}^2\Theta}{\text{d}\theta^2}
-\cos\theta\frac{\text{d}\Theta}{\text{d}\theta},
\end{eqnarray}
which integrate as
\begin{subequations}\label{eq:RTP}
\begin{eqnarray}
\Phi(\phi) &=& A_1\cos\phi+A_2\sin\phi,\\
\Theta(\theta) &=& A_3\cos\theta, \\
R(r) &=& \left\{
\begin{array}{l l}
B_{-}\sqrt{1-kr^2}, & k\ne0,\\ \\
\frac12\alpha{r}^2, & k=0,
\end{array}\right.
\end{eqnarray}
\end{subequations}
modulo the previously mentioned freedom in the election of the
above functions. Here, $A_i$ (with $i=1,2,3$), $\alpha$ and
$B_{-}$ are integration constants.

Using now the combination
\begin{equation}\label{eq:T_te^te-T_t^t}
3a^4\sigma^3\left(T_{~\theta}^{\text{s}~\theta}
-T_{~t}^{\text{s}~t}\right) =\left\{
\begin{array}{l l}
\frac{\text{d}^2}{\text{d}\tau^2}T+kT=0,
& k\ne0,\\ \\
\frac{\text{d}^2}{\text{d}\tau^2}T +\alpha=0, & k=0,
\end{array}\right.
\end{equation}
we have that the dependence on the conformal time is
\begin{equation}\label{eq:T(tau)}
T(\tau) =\left\{
\begin{array}{l l}
-\frac{A_0}{\sqrt{k}}\sin\left(\sqrt{k}\tau\right)+B_{+}\cos\left(\sqrt{k}\tau\right),
& k\ne0,\\ \\
-\frac12\alpha\tau^2-A_0\tau+\sigma_0, & k=0,
\end{array}\right.
\end{equation}
where $A_0$, $B_{+}$ and $\sigma_0$ are integration constants.
The above temporal dependences of the stealth make clear why
the use of the conformal time is essential; using the comoving
time $t=\int{d}\tau{a(\tau)}$ the solution is expressed in
terms of quadratures of the scale factor!

Only one equation remains in our study, and is independent of
the value of the curvature if one makes the following
redefinitions of the integration constants of the curved cases
$B_{\pm}=\sigma_0/2\pm\alpha/k$,
\begin{equation}\label{eq:lambda}
-2a^4\sigma^4T_{~t}^{\text{s}~t}=
\lambda+{A_0}^2-\vec{A}^2+2\alpha\sigma_0=0,
\end{equation}
which shows that one of the integration constants is fixed in
terms of the coupling constant of the conformal potential and
the other integration constants.

Therefore, for any standard cosmology, independently of its
spatial topology and matter content, there exists a conformal
stealth generally described by Eq.~(\ref{eq:Psi2sigma}) with
$\sigma$ written as in Eq.~(\ref{eq:sigma_s}), and functions
$R$, $\Theta$, $\Phi$ and $T$ given by Eqs.~(\ref{eq:RTP}) and
(\ref{eq:T(tau)}). Finally, one of the involved integration
constants is not independent and is determined by the coupling
constant $\lambda$ and the remaining integration constants from
Eqs.~(\ref{eq:lambda}).

Next, we analyze separately the explicit form of the conformal
stealth for flat and curved universes, generalize these results
to any number of dimensions and study how many integration
constants can be eliminated using spacetime symmetries in each
case.

\section{Flat Universes\label{sec:flat}}

For universes with flat spatial topology, $k=0$, the FRW metric
(\ref{eq:FRW}) becomes manifestly conformally flat
\begin{equation}\label{eq:FRWk=0}
ds^2=a(\tau)^2\eta_{\mu\nu}dx^{\mu}dx^\nu.
\end{equation}
Therefore, in this case we indeed expect the expression for the
stealth (\ref{eq:Psi2sigma}) to be exactly a conformal
transformation of the stealth for Minkowski flat spacetime
\begin{equation}\label{eq:k=0S_CT}
\Psi=\frac1a\Psi_{\text{flat}}.
\end{equation}
To check this, we combine the results for $k=0$ of
Eqs.~(\ref{eq:sigma_s}), (\ref{eq:RTP}), and (\ref{eq:T(tau)})
of the previous section to obtain
\begin{equation}\label{eq:sigmak0}
\sigma(x^\mu)=\frac{\alpha}2x_{\mu}x^{\mu}+A_{\mu}x^{\mu}+\sigma_{0},
\end{equation}
where we raise and lower indices with the flat metric. This is
precisely the result for flat spacetime found in
Ref.~\cite{AyonBeato:2005tu}.

The above results are easy to generalize to any number of
dimensions using the following recipe. The metric and the
auxiliar function $\sigma$ are still given by expressions
(\ref{eq:FRWk=0}) and (\ref{eq:sigmak0}), but with $\mu$ and
$\nu$ now running from $0$ to $D-1$. The conformal stealth must
be written now as
\begin{equation}\label{eq:Psi2sigmaD}
\Psi=\frac1{(a\sigma)^{(D-2)/2}}\, .
\end{equation}
Then, in the D-dimensional version of action
(\ref{eq:action+s}), the conformal coupling must be generalized
to
\begin{equation}\label{eq:ConfCopD}
\frac{1}{6} \quad\longrightarrow\quad \frac{D-2}{4(D-1)},
\end{equation}
and similarly for the conformal potential
\begin{equation}\label{eq:ConfPotD}
\frac12\,\lambda\,\Psi^4\quad\longrightarrow\quad\frac{(D-2)^2}{8}\,\lambda\,\Psi^{2D/(D-2)}.
\end{equation}

The relation between the coupling constant $\lambda$ and the
integration constants is again determined by
Eq.~(\ref{eq:lambda}), where vectorial quantities have now
$D-1$ components.

Finally, it is worthy to emphasize an important difference
between the conformally related versions of the stealths in
Minkowski spacetime and flat universes. In Minkowski spacetime
due to translational invariance it is possible to fix the
constants $A_\mu$ in expression (\ref{eq:sigmak0}) to zero,
hence, to describe the conformal stealth only one integration
constant is required and the related solution is manifestly
Lorentz invariant \cite{AyonBeato:2005tu}. Flat universes
(\ref{eq:FRWk=0}) have translation invariance only along
spatial directions, which allows to choose as vanishing only
the spatial constants $A_i$ in (\ref{eq:sigmak0}).
Consequently, the conformal stealth of flat universes allows
for two integration constants and is only manifestly isotropic.

\section{Curved Universes\label{sec:curved}}

For universes with curved spatial topology, $k=\pm1$, the FRW
metric (\ref{eq:FRW}) can be rewritten as
\begin{equation}\label{eq:FRWk=+-1}
ds^2=a(\tau)^2\left(-d\tau^2+d\vec{x}^2
+\frac{k\,(\vec{x}{\cdot}d\vec{x})^2}{1-k\,\vec{x}^2}\right),
\end{equation}
using standard Euclidean coordinates. Combining again
Eqs.~(\ref{eq:sigma_s}), (\ref{eq:RTP}), and (\ref{eq:T(tau)})
the function $\sigma$ characterizing the corresponding stealth
is given by
\begin{eqnarray}\label{eq:sigmak+-1}
\sigma(x^\mu)&=&-\frac{A_0}{\sqrt{k}}
\sin\left(\sqrt{k}\tau\right)
+\left(\frac{\sigma_0}2+\frac{\alpha}{k}\right)
\cos\left(\sqrt{k}\tau\right)\nonumber\\
&&{}+\vec{A}\cdot\vec{x}
+\left(\frac{\sigma_0}2-\frac{\alpha}{k}\right)
\sqrt{1-k\,\vec{x}^2},
\end{eqnarray}
where we have used the redefinition on the integration
constants giving the universal relation (\ref{eq:lambda}) to
the coupling constant $\lambda$. Notice that as a byproduct
these redefinitions allows to recover consistently the flat
case (\ref{eq:sigmak0}) by taking the limit $k\rightarrow0$.

Expressions (\ref{eq:FRWk=+-1}) and (\ref{eq:sigmak+-1}) allow
obvious generalizations to higher dimension, which define a
higher-dimensional conformal stealth also for the curved cases.
It is only needed to follow the outlines given in the previous
section.

Here, the standard spatial translation invariance is broken due
to the presence of spatial curvature, however, a generalization
of spatial translations still remains as symmetry. These
quasitranslations \cite{Weinberg:1972} can be understood as
follows. As is well-known, the constant curvature spatial
sections can be isometrically embedded in a flat space with one
extra dimension whose rotations induce the constant curvature
isometries. The spatial coordinates of metric
(\ref{eq:FRWk=+-1}) are just the embedding coordinates, and
rotations along the planes orthogonal to the extra dimension in
the ambient space induce just the isotropy of metric
(\ref{eq:FRWk=+-1}). Moreover, rotations along the planes
formed with the extra dimension induce the quasitranslations,
which have the following explicit form \cite{Weinberg:1972}
\begin{equation}\label{eq:quasiT}
\vec{x}\mapsto\vec{x}+\vec{a}\left(\sqrt{1-k\,\vec{x}^2}
-\frac{\left(1-\sqrt{1-k\,\vec{a}^2}\right)}{\vec{a}^2}
\vec{a}{\cdot}\vec{x}\right).
\end{equation}
For $k=0$ they become just standard translations. For any
curvature, these transformations map the origin $\vec{x}=0$ to
any arbitrary point $\vec{x}=\vec{a}$, what is an explicit
realization of the homogeneous character of spacetime
(\ref{eq:FRWk=+-1}). Under quasitranslations, metric
(\ref{eq:FRWk=+-1}) is invariant but the stealth is just form
invariant, i.e.\ its local dependence after the transformation
is the same but with transformed integration constants. This
allows to choose specific values for the quasitranslations
parameters $\vec{a}$ such that the transformed integration
constants $\vec{A}$ acquire vanishing values. This is achieved
choosing the parameters as
\begin{equation}\label{eq:a2A=0}
\vec{a}=\frac{\vec{A}}{\sqrt{\left(\alpha
-\frac{k}{2}\sigma_0\right)^2+k\vec{A}^2}}.
\end{equation}
Notice, that consequently, in the limit $k\rightarrow0$ these
become just the standard translations annihilating the vector
$\vec{A}$ in the flat case. Due to the above argument, after
considering the relation to the coupling constant
(\ref{eq:lambda}), the conformal stealth of curved universes
has also only two integration constants and is manifestly
isotropic as in the case of flat universes.

\subsection{Conformal transformation from Minkowski spacetime}

According to our discussion of Sec.~\ref{sec:SfS}, in a proper
limit, it should be possible to reduce expression
(\ref{eq:sigmak+-1}) to the result obtained by conformally
mapping the stealth from Minkowski spacetime. We will take that
into account to crosscheck (\ref{eq:sigmak+-1}) and also to
gain a deeper insight into the role of the symmetries in
stealth configurations. With these aims we use the following
map between Minkowski and FRW spacetimes \cite{Lightman:1975}
\begin{subequations}\label{eq:var_chang}
\begin{eqnarray}
\tau &=& \frac{1}{\sqrt{k}}\arctan\left(\frac{\sqrt{k}t}{1-\frac{k}{4}(t^2-\rho^2)}\right),\\
r &=& \frac{1}{\sqrt{k}}\sin\left[\arctan\left(\frac{\sqrt{k}\rho}{1+\frac{k}{4}(t^2-\rho^2)}\right)\right],
\end{eqnarray}
\end{subequations}
with inverse given by
\begin{subequations}\label{eq:inv_var_chang}
\begin{eqnarray}
t &=& \frac{2\sin(\sqrt{k}\tau)}{\sqrt{k}\left(\cos(\sqrt{k}\tau)+\sqrt{1-kr^2}\right)},\\
\rho &=& \frac{2r}{\cos(\sqrt{k}\tau)+\sqrt{1-kr^2}}\, .
\end{eqnarray}
\end{subequations}
It is important to notice here that these expressions provide
only a local map. For $k=1$ the whole Minkowski spacetime is
mapped into just a patch of FRW, while conversely, for $k=-1$,
just a patch of the Minkowski spacetime is mapped into the
whole FRW.

We develop the conformal argument in any dimension, since there
is nothing particular in the tetra-dimensional case. This way,
the Friedman-Robertson-Walker metric in D dimensions can be
written as \cite{Lightman:1975}
\begin{eqnarray}\label{eq:expl_conf}
ds^2_{\text{FRW}}&=&a(\tau)^2\left(-d\tau^2+\frac{dr^2}{1-kr^2}+r^2d\Omega^2_{D-2}\right)\nonumber\\
&=&\frac{a\big(\tau(t,\rho)\big)^2\left(-dt^2+d\rho^2+\rho^2d\Omega^2_{D-2}\right)}
{\left[\frac{k}{4}(\rho+t)^2+1\right]\left[\frac{k}{4}(\rho-t)^2+1\right]}\nonumber\\ \nonumber\\
&=&\Omega^2ds^2_{\text{M}}.
\end{eqnarray}
Consequently, the conformal transformation for the stealth will
be
\begin{eqnarray}
\Psi_{\text{FRW}} &=& \frac{1}{\Omega^{(D-2)/2}} \Psi_{\text{M}}\nonumber\\
&=& \frac1{(\Omega \sigma_{\text{M}})^{(D-2)/2}}\, ,
\end{eqnarray}
where the conformal factor $\Omega$ can be drawn from
(\ref{eq:expl_conf}) and the auxiliary function
$\sigma_{\text{M}}$ of Minkowski spacetime (\ref{eq:sigmak0})
is rewritten now as
\begin{equation}\label{eq:Omega_sigMink}
\sigma_{\text{M}}=\frac{\alpha}{2}(-t^2+\rho^2)-A_0t+\rho A_m\pi^m+\sigma_0,
\end{equation}
where $\pi^m$ are the polar coordinates of the $S^{D-2}$ unit
sphere. Using the inverse transformations
(\ref{eq:inv_var_chang}) the conformal factor reduces to
\[
\Omega = \frac{a(\tau)}{2}\left[\cos(\sqrt{k}\tau)+\sqrt{1-kr^2}\right],
\]
while for $\sigma_{\text{M}}$ we obtain
\begin{eqnarray}
\sigma_{\text{M}}=2\Biggl(&-&\frac{\alpha}{k}\frac{\cos(\sqrt{k}\tau)-\sqrt{1-kr^2}}
                                    {\cos(\sqrt{k}\tau)+\sqrt{1-kr^2}}\nonumber\\
&-&\frac{A_0}{\sqrt{k}}\frac{\sin(\sqrt{k}\tau)}{\cos(\sqrt{k}\tau)+\sqrt{1-kr^2}}\nonumber\\
&+&\frac{r A_m \pi^m}{\cos(\sqrt{k}\tau)+\sqrt{1-kr^2}}+\frac{\sigma_0}{2}\Biggr).
\end{eqnarray}
Combining these last two expressions lead us finally to
Eq.~(\ref{eq:sigmak+-1}), i.e.\
$\Omega\sigma_{\text{M}}=a\sigma_{\text{FRW}}$ and the local
expression for the stealth obtained from the conformal
transformation coincides exactly with the general previously
founded solution.

\section{Conclusions}

We have proved that any homogeneous and isotropic universe,
independently of its spatial topology and matter content,
allows for the existence of a conformal stealth. Surprisingly,
though the stealth is isotropic, it is not homogeneous.
Nevertheless, its presence leaves no trace in the cosmological
evolution of the given universe. Additionally, we have shown
that these results are not exclusive of our four-dimensional
universe, but are also valid for higher-dimensional
generalizations of the FRW spacetime. We prove all the above by
solving explicitly the related constraints, but we also discuss
to some extension the local conformal arguments that
alternatively allow to build such configurations from the ones
of Minkowski spacetime. After making the explicit construction
we show that the potential problems that are known to occur in
other contexts, due to the local nature of these arguments, are
not present in the cosmological framework. However, due to the
fact that the involved conformal factors break in general the
maximally symmetric character of Minkowski spacetime, the
resulting conformally generated stealth configurations allow
more integration constants than its seeds from Minkowski
spacetime. For FRW spacetimes the conformal factors break
time-translation invariance and as consequence its stealths
allow one additional integration constant in comparison to the
single one allowed by its conformally related cousins of
Minkowski spacetime.

Last, but not least, it is important to understand the observational 
consequences of the existence of cosmological stealths. In this 
sense, we note that its fluctuations are not expected to be stealth 
themselves. This way, the corresponding stealth perturbations may have 
an imprint on the spectra of the cosmic microwave background radiation 
as well as in the statistics of the cosmological large scale 
structures. Exploring these consequences is the subject of our current research program.

\begin{acknowledgments}
The authors thank enlightening discussions with Mokthar
Hassa\"{i}ne, Tonatiuh Matos, Bogdan Mielnik and Massimo
Porrati. This work has been partially supported by CONACyT
grants 175993 and 178346, PROMEP grant PROMEP/103.5/10/4948,
and FONDECYT grants 1090368, 1130423, 11090281 and 1121031. One of the
authors (CAT-E) would like to thank the Department of Physics,
CINVESTAV, Mexico, for kind hospitality extended to him during
part of this investigation.
\end{acknowledgments}


\end{document}